\documentclass[aps,prd,%preprint,
showpage,nobibnotes,groupaddress,amssymb,amsmath,nofootinbib,floatfix]{revtex4-2}

\usepackage{graphicx,amsfonts}
\usepackage{array,booktabs}
\usepackage{epsfig}
\usepackage{dcolumn}
\usepackage{bm}
\usepackage{siunitx}
\usepackage[hidelinks]{hyperref}

\begin{document}

\title{Can scalars matter? A vector bilepton phenomenologic example}

\author{Mario W. Barela}
\email{mario.barela@unesp.br}
\affiliation{
Instituto de F\'\i sica Te\'orica, Universidade Estadual Paulista, \\
R. Dr. Bento Teobaldo Ferraz 271, Barra Funda\\ S\~ao Paulo - SP, 01140-070,
Brazil}

\begin{abstract}
Expecting scalar contributions to be less important to a given observable signal than vector ones, in usual scenarios, is a natural intuition. Such assertion should hold for a great part of physically relevant parameter space within most models of interest. In this work, besides stressing that the proposition is not general, we show how the alternative possibility may give light to interesting phenomena.
\end{abstract}

\maketitle

\section{Introduction}
\label{sec:intro}

To accommodate a major part of the current observed data and theoretical necessities within a fundamental model is an enormous challenge. A common price to pay when pursuing this goal is the introduction of extra particles, which, by themselves, are usually not desirable. Exotic scalar particles, for instance, are crucial for the fitting of known fermion masses and may be important in the precision fitting of loop observables.

Because of this reality, ultraviolete complete models proposed as extensions or alternatives to the SM, supposed to deal with the standard problems of current particle physics, tend to contain a number of exotic particles which should be skeptically assessed phenomenologically. This, however, is in most cases impractical because of the great number of free parameters introduced together with the new degrees of freedom. 

Among the simplifications imposed in order to allow for phenomenology of these models to be developed (which include benchmarks motivated by, e.g., perturbativity and naturalness) is deeming scalar contributions insignificant or unimportant altogether. Since there is less angular momenta configurations allowed in scalar-mediated processes, this is justified and even expected for a great part of parameter space in most specific cases -- however, as will be showed in this note, not a quantitatively safe assumption in the general case.

We choose a single observable -- namely the muon branching ratio of the tree-level $\mu^+ \to e^+ e^+ e^-$ process -- and show how the existence of a neutral scalar can considerably \textit{relieve} the exclusion contour on the mass of a doubly charged vector bilepton. This scenario is pertinent, for instance, within a 331 model \cite{Pisano:1991ee,Frampton:1992wt} or a $SU(15)$ GUT model \cite{Frampton:1989fu}, which are the known ultraviolete complete theories to contain a doubly charged vector bilepton.

\section{Interactions}
\label{sec:int}

Our mindset is focused in showing that the proposition above -- regarding the importance of scalar contributions -- is not generally safe by counterexample. Our tool will be the model independent constraints implied by the flavour violating decay $\mu^+ \to e^+ e^+ e^-$ on the mass of a doubly charged vector bilepton $U^{++}$. The relevant $U^{++}$-lepton-lepton interaction may be generally parametrized like

 \begin{equation}\label{eq:Uint}
 \begin{split}
\mathcal{L}_{U\ell\ell}&=\sum_{b> a} \frac{g_{2L}}{\sqrt{2}} \left\{ U_\mu^{++} \bar{\ell_a^c}\gamma^\mu [P_L (V_U)_{ab}-P_R (V_U)_{ba}] \ell_b + U_\mu^{--}\bar{\ell_a}\gamma^\mu [P_L (V_U^\dagger)_{ab}-P_R (V_U^\dagger)_{ba}] \ell_b^c\right\} + \\
&+ \sum_{a} \frac{g_{2L}}{\sqrt{2}} \left\{ U_\mu^{++} \bar{\ell_a^c}\gamma^\mu [P_L (V_U)_{aa}] \ell_a + U_\mu^{--}\bar{\ell_a}\gamma^\mu [P_L (V_U^\dagger)_{aa}] \ell_a^c\right\},
\end{split}
\end{equation}
where $P_{L,R}$ is the left(right)-handed projector and $V_U$ is a unitary matrix, with $a=e,\mu,\tau$. 

The second particle considered is a flavor violation-mediating neutral scalar $s$. The interaction is written as

\begin{equation}
\mathcal{L}_{s\ell\ell}=-\sum_{a,b}g_{sL}\bar{\ell_a}\left[(\mathcal{O}_s)_{ab} P_L+(\mathcal{O}^\dagger_s)_{ab} P_R\right] \ell_b s,
\end{equation} 
where $\mathcal{O}_s$ is an arbitrary matrix related to the Yukawa couplings in a specific model. Also, $g_s$ is, in practice, absorbed into $\mathcal{O}_s$.

So far, the entire construction is general and model independent. As a simplification, we consider the matrix $V_U$ to be orthogonal, ignoring the 6 phases that a unitary matrix could entail. We write it as

\begin{equation}
V_U = \begin{pmatrix}
\cos(\psi)\cos(\phi)-\cos(\theta)\sin(\phi)\sin(\psi) & \cos(\psi)\sin(\phi)+\cos(\theta)\cos(\phi)\sin(\psi) & \sin(\theta)\sin(\psi) \\
-\sin(\psi)\cos(\phi)-\cos(\theta)\sin(\phi)\cos(\psi) & -\sin(\psi)\sin(\phi)+\cos(\theta)\cos(\phi)\cos(\psi) & \sin(\theta)\cos(\psi) \\
\sin(\theta)\sin(\phi) & -\sin(\theta)\cos(\phi) & \cos(\theta)
\end{pmatrix}.
\end{equation}

Lastly, to fix the range of both matrix elements we demand that every modulus obey $|(V_U)_{ab}|,|(\mathcal{O}_s)_{ab}|>10^{-3}$, so that there can be no large unnatural hierarchy.

\section{Results}
\label{sec:res}

\begin{figure}[t!]
{\centering
\hspace*{2cm} \includegraphics[width=0.65\linewidth]{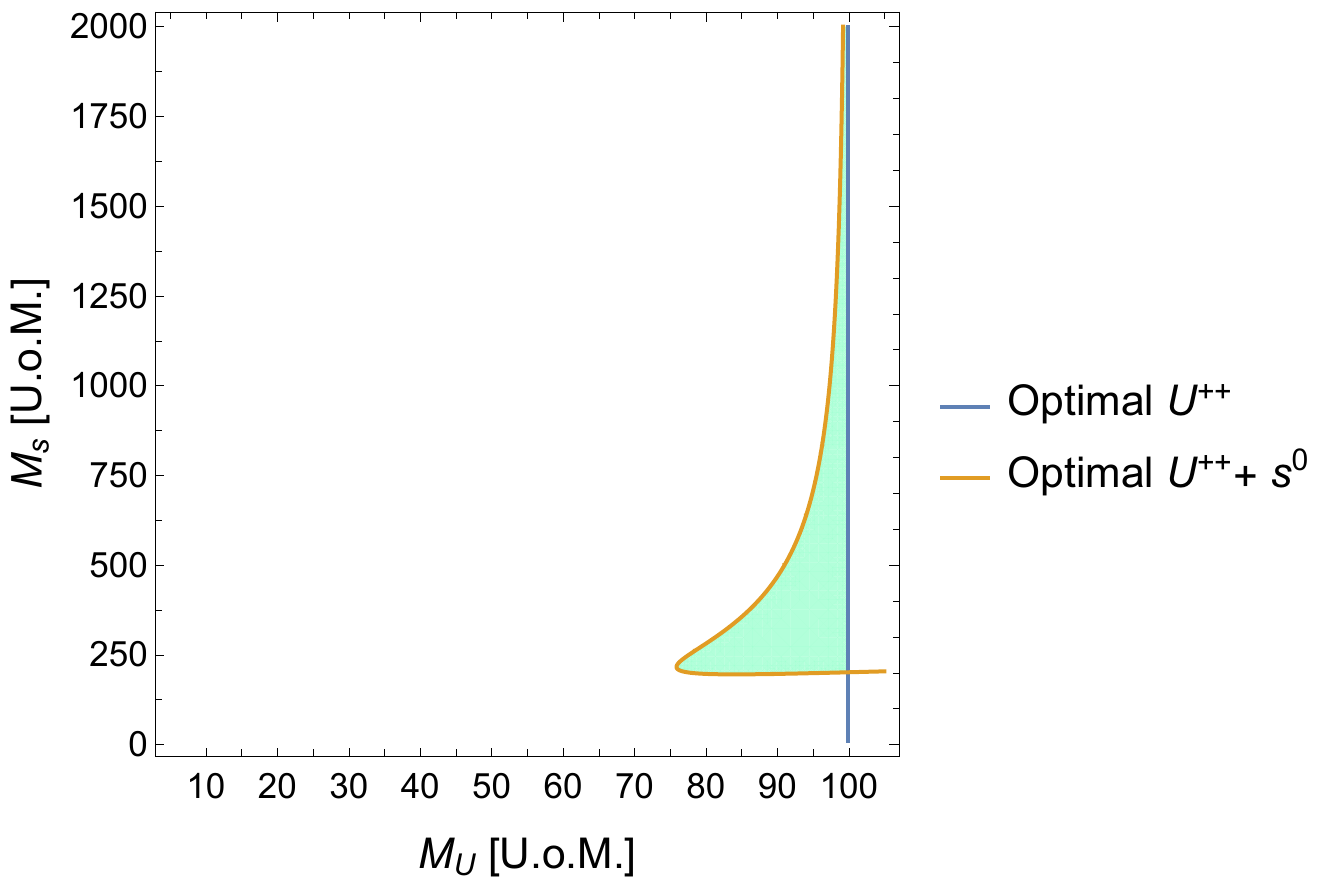}
\caption{Optimal exclusion contours in the scenario with only the $U^{++}$-particle and in the one with it together with the scalar $s$. The axes are presented in arbitrary Units of Mass, scaled in such a way that the pure $U^{++}$ contour is set at 100.}
\label{fig:Contours}
}
\end{figure}

We will search for \textit{optimal} sets of the 6 free parameters, as described in the last section, which saturate the experimental limit $\text{Br}\left(\mu^+  \to  e^+ e^- e^+\right)<10^{-12}$ \cite{Zyla:2020zbs}. A solution is optimal if it allows for the weakest bounds (at least in a single point) on the mass of the $U^{++}$, both in a scenario where only it is present and in one where the scalar particle also exists. The solution appears on Table \ref{Tab:sol} and the results are presented on Figure \ref{fig:Contours}.

\begin{table}[b!]
\caption{Parameter values that optimize the exclusion contours.}\label{Tab:sol}
\begin{tabular}{c@{\hskip 1.5em}c@{\hskip 38pt}c@{\hskip 1.5em}r}
\toprule
\multicolumn{2}{c}{$U$}                             &  \multicolumn{2}{c}{$U-s$}                                 \\ \midrule
$\phi$               &   $4.71139$                  &  $\phi$                &   \multicolumn{1}{c}{$4.71338$}   \\
$\psi$               &   $1.56980$                  &  $\psi$                &   \multicolumn{1}{c}{$1.56980$}   \\
$\theta$             &   $1.57180$                  &  $\theta$              &   \multicolumn{1}{c}{$1.56980$}   \\
                     &                              &  $\mathcal{O}_{s11}$   &   $-1.01913 \times 10^{-3}$       \\
                     &                              &  $\mathcal{O}_{s12}$   &   $2.11077 \times 10^{-3}$        \\
                     &                              &  $\mathcal{O}_{s21}$   &   $2.11077 \times 10^{-3}$        \\ \bottomrule
\end{tabular}
\end{table}

We observe that the introduction of a scalar particle may, in the optimal sector of parameter space, cause a relevant new area in the plane $M_U \times M_s$ to be allowed, relieving the bounds on the vector bilepton $U^{++}$ by as much as 25\% when the $s$ mass is around 2.2 times bigger than $M_U$. 

The results are presented in arbitrary units, and the ones in natural units are obtained by rescaling through multiplication by $1.1$ -- this means that the considered process alone is not efficient, in general, in constraining the masses of our particles, maintaining masses of the order of $\SI{100}{GeV}$ still possible. We stress that this bounds are not meant to be useful in practice and, to obtain appropriate ones, more processes should be considered simultaneously -- this, however, does not factor into our discussion. 

Finally, it is important to note that, analyzing the solution in Table \ref{Tab:sol}, we may verify that the vector contribution is indeed dominant since its optimal parameters remain practically unchanged to 5 decimal places between the two scenarios, and the optimal scalar mixing matrix elements are of the lowest allowed order of magnitude, $10^{-3}$. However, besides not changing our conclusions that scalars may matter and cannot be skeptically ignored, this could be a consequence of the specific construction of a vector bilepton, which is theoretically expected to mix leptons through a unitary matrix; and of the imposition on the modulus of every matrix element.  

\section{Conclusions}
\label{sec:con}

It is operationally very difficult to test sectors of a model with a number of relevant free parameters greater than a handful. In fact, even if that was strictly possible, the resulting set of data would be of little use in many cases, since it tends to be hard to extract or interpret correlations. In consequence, it is common, while doing phenomenology, to isolate the supposedly minimal necessary part of the model. While natural and acceptable, oftentimes in order to do so, the imposition of non-general or unverifiable assumptions or simplifications is inevitable.

As a representation, among several works which treat hadronic processes at the LHC with a 4-lepton end state \cite{Meirose:2011cs,Dion:1998pw,Nepomuceno:2016jyr}, we have, for instance, Ref.~\cite{Barela:2019pmo} which concerns itself only with the bilepton contribution to a hadronic process, to be able to assess it in a skeptical fashion, while Ref. \cite{Coriano:2018khp} consider more particles simultaneously, but choosing arbitrary benchmark points on parameter space.

In this note we call attention to the fact that even if valid for a greater section of parameter space, assumptions regarding the unimportance of scalar contributions with respect to vector ones must be taken responsibly since they tend to not be general and to overshadow interesting phenomena. Specifically, we show how the influence of a scalar may considerably change the theoretical consequences of the phenomenology of a single observable, relieving the optimal bounds on the mass of a doubly charged vector bilepton by as much as 25\%, which urges for studies that focus in reconsidering the importance of non-dominant contributions to the fitting of signals through, for instance, reinterpretation of LHC results \cite{LHCReinterpretationForum:2020xtr}.

\section*{ACKNOWLEDGMENTS}
MB would like to thank CNPq for the financial support.

\end{document}